\documentclass[aps,prl,reprint,superscriptaddress,showpacs]{revtex4-1}

\usepackage[utf8]{inputenc}

\usepackage{amsmath}
\usepackage{graphicx}

\newcommand{\braket}[3]{\ensuremath{\left\langle #1 \left| #2 \right| #3 \right\rangle}}

\begin{document}
\title{Charge doping versus impurity scattering in chemically substituted iron-pnictides}
\author{Alexander Herbig}
\affiliation{Institut f\"ur Festk\"orperphysik, Karlsruher Institut f\"ur Technologie, 76021 Karlsruhe, Germany}
\author{Rolf Heid}
\affiliation{Institut f\"ur Festk\"orperphysik, Karlsruher Institut f\"ur Technologie, 76021 Karlsruhe, Germany}
\author{J\"org Schmalian}
\affiliation{Institut f\"ur Theorie der Kondensierten Materie, Karlsruher Institut f\"ur Technologie, 76131 Karlsruhe, Germany}
\affiliation{Institut f\"ur Festk\"orperphysik, Karlsruher Institut f\"ur Technologie, 76021 Karlsruhe, Germany}

\date{\today}

\begin{abstract}
 To reveal the relative importance of charge doping and defect scattering in substitutionally modified 122 iron pnictides, we perform a systematic first principles study on selected bands at the Fermi level. Disorder effects
 are induced by various substitutions using an orbital based coherent potential approximation (CPA). Pronounced level shifts of individual bands suggest that transition metal substitutions introduce mobile charge carriers into 
 the system. However, important deviations from such a rigid band scenario as well as spectral broadenings due to impurity scattering correlate with the band character. Finally a $T$-matrix analysis exhibits a larger intraband 
 than interband scattering consistent with an $s^{+-}$ pairing state. Comparing different substitutions reveals an increase of pair-breaking along the transition metal series.
\end{abstract}


\maketitle

\begin{figure}[t]
  \includegraphics[width=1.0\columnwidth]{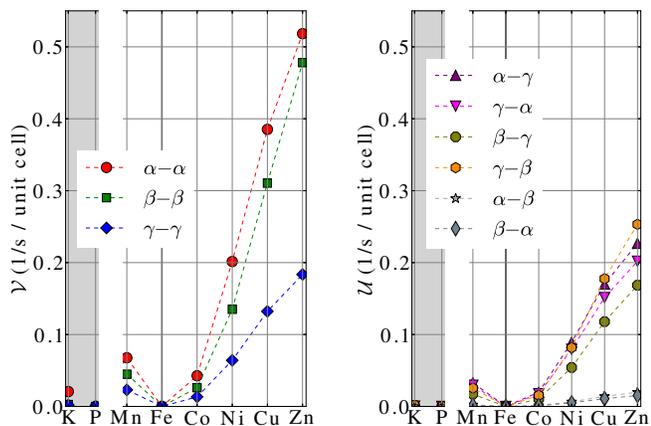}
  \caption{(color online) Intraband $\mathcal{V}$ and interband $\mathcal{U}$ scattering rates obtained from the single-impurity $T$-matrix in dilute limit}
   \label{fig:tmat}
\end{figure}

Chemical substitution is an important tuning parameter which governs the phase diagram and thus the onset of superconductivity in the various iron pnictide superconductors. Despite intensive experimental and theoretical work on
a variety of different families of these compounds for more than half a decade by now, the role of substitutional disorder is still under debate. In particular the substitution of Fe by other 3$d$ transition metals (TM) in the 
122-family such as BaFe$_2$As$_2$ has been discussed controversially. Macroscopic measurements over a range of compositions \cite{canfield2009} suggest that the number of extra $d$ electrons at the TM site is the 
decisive quantity which determines the shape of the superconducting dome for TM\,$\in\{$Co,Ni$\}$. Such a rigid-band-shift scenario is also compatible with changes of Fermi surfaces as observed in angular resolved photoemission
spectroscopy (ARPES)\cite{liu2010,neupane2011}. On the other hand, X-ray absorption measurements (NEXAFS) \cite{merz2012} see at best a small change of valence at the Fe atom induced by TM substitution, which challenges the view of a 
rigid band shift. This finding is also consistent with the dependence of the N\'eel temperature on chemical substitution reported in \cite{canfield2009}. From the viewpoint of electronic structure calculations this dichotomy 
between localized extra electrons and doping into conduction bands has been first addressed by supercell calculations \cite{wadati2010,berlijn2012}.  More recently, effective medium approaches, which can handle arbitrary 
impurity concentrations, have been used to study the effect of substitutional disorder on bandstructure and Fermi surface topology \cite{khan2014,derondeau2014}. Nevertheless a systematic first principles investigation of 
different substitutions on the behavior of electronic quasiparticles is still lacking.\\
\indent A further aspect of substitutional disorder is its impact on the superconducting state. It is widely accepted that these systems are unconventional superconductors where impurity scattering is important for Cooper 
pair-breaking \cite{balian1963,annett1991,golubov1997,mackenzie2003,balatsky2006,alloul2009,kogan2009,hirschfeld2011}. For example the popular $s^{+-}$ pairing state would be more susceptible to interband scattering \cite{vorontsov2009} than the more conventional $s^{++}$ state. Thus, knowledge of the 
band-resolved scattering rates induced by the different substituents can shed light on the symmetry of possible superconducting gap functions.\\
\indent In this Letter we performed electronic structure calculations of substitutionally disordered iron based systems to address these topics. We focus on disorder-induced level shifts and spectral broadenings due to impurity 
scattering or, equivalently, self-energy effects on individual hole and electron bands at the Fermi level. We systematically investigated the Ba(Fe$_{1-x}$TM$_x$)$_2$As$_2$ -series where TM\,$\in\{$Mn, Co, Ni, Cu, Zn$\}$ as well as 
the Ba$_{1-x}$K$_x$Fe$_2$As$_2$ and the BaFe$_2$(As$_{1-x}$P$_x$)$_2$ systems at impurity concentrations $x<0.1$. Additionally we analyzed the impact of these substitutions on intra- and interband 
scattering in the limit of small impurity concentrations.\\
\indent Our major findings are: (i) impurity substitution leads to both level shifts and spectral broadening, i.e. TM substitutions simultaneously supply mobile carriers and act as scattering centers; (ii) the magnitude of 
level shifts and degree of spectral broadenings depend sensitively on the orbital composition of the respective Fermi surface sheet; (iii) a stronger intraband scattering on hole bands dominating the transport properties and a 
weaker interband scattering between electron and hole bands render the $s^{+-}$ pairing state \cite{zhang2009,sknepnek2009} comparatively robust.\\
\indent In our non spin-polarized electronic structure calculations the substitutional disorder was treated within Blackman, Esterling and Berk's \cite{blackman1971} extension of the coherent potential approximation 
\cite{taylor1967,soven1967} (BEB-CPA) which is an effective medium method dedicated to the treatment of arbitrary impurity concentrations. Unlike the conventional CPA, which only can handle onsite disorder, the BEB-CPA 
additionally allows to incorporate off-diagonal disorder effects on the level of disordered hopping terms. Following earlier work \cite{koepernik1997}, the charge self-consistent BEB-CPA was implemented using a non-orthogonal 
set of atom-centered basis functions which were obtained from fits to \emph{ab-initio} bandstructures of the parent compound BaFe$_2$As$_2$ and respective substitutional end-member, for example BaTM$_2$As$_2$, by means of density 
functional theory (DFT) which also provided the required potentials to calculate the BEB-CPA Hamiltonian. We used nine local basis functions up to an orbital angular momentum of $l=2$ for each atomic species.  All calculations 
were done in the tetragonal high-temperature phase of the stoichiometric BaFe$_2$As$_2$ with the structural parameters given in \cite{Drotziger2010} using the 2-Fe and 2-As unit cell. The crystal structure was not changed with 
impurity concentration $x$ in order to separate disorder from structural effects. The DFT calculations were performed within the mixed-basis-pseudopotential approach (MBPP code) \cite{meyer1997,louie1979} applying the local 
density approximation (LDA) \cite{perdewWang1992}. The Brillouin zone integration was performed on a 8x8x4 Monkhorst-Pack \cite{monkhorst1976} $\mathbf{k}$-mesh. We used norm-conserving pseudoptentials constructed after Vanderbilt 
\cite{vanderbilt1985} together with local $d$-type functions for Fe or TM respectively and plane waves up to a cutoff energy of 22 Ry.\\
\begin{figure}[t]
\includegraphics[width=1.0\columnwidth]{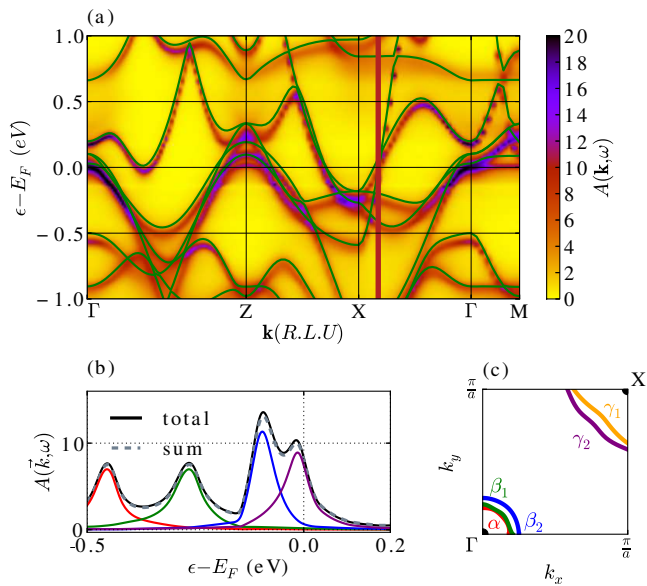}
\caption{(color online) (a) Bloch spectral function of disordered Ba(Fe$_{0.9}$Ni$_{0.1}$)$_2$As$_2$ together with band structure of the parent compound (green solid lines); (b) Bloch spectral function of 
Ba(Fe$_{0.9}$Ni$_{0.1}$)$_2$As$_2$ at a single $\mathbf{k}$-point [vertical red line in (a)] together with projections (red, green, blue, purple) on bands of the parent compound and sum over all bands; (c) Fermi surface of 
the parent compound in the ($\mathbf{k}_x,\mathbf{k}_y$) plane at $\mathbf{k}_z=0$}
\label{fig:bstruc_spec_fsurf}
\end{figure}
\indent For TM substitutions we find similar modifications of the density of states (DOS) as in previous supercell calculations \cite{berlijn2012}. In particular, only at Zn substitution a localized state forms at
7eV below the Fermi level. To obtain a deeper insight into effects of disorder on selected bands near the Fermi level we considered the Bloch spectral function
\[
 A(\mathbf{k},\omega) = -\frac{1}{\pi} \mathrm{Im}\mathrm{Tr}\left[S(\mathbf{k})\Gamma(\mathbf{k},\omega)\right]
\]
where $\Gamma$ is the BEB-CPA effective-medium Green's function, $S$ is the overlap matrix and the trace is taken over local basis indices. In Fig.\,\ref{fig:bstruc_spec_fsurf}(a) $A(\mathbf{k},\omega)$ is plotted in false color 
over energy along a path in $\mathbf{k}$-space for disordered Ba(Fe$_{0.9}$Ni$_{0.1}$)$_2$As$_2$ together with the bandstructure of the parent compound. It contains all information about the effects of disorder on the 
bandstructure, in particular shifts with respect to the parent compound and $\mathbf{k}$-dependent spectral widths in energy. While this is useful for an overview it is not well suited for a systematic analysis of the behavior 
of individual bands due to its multi-peak structure. An example is given in Fig.\,\ref{fig:bstruc_spec_fsurf}(b) where the solid black line shows the spectral function at a fixed $\mathbf{k}$-point next to X indicated by the 
vertical red line in Fig.\,\ref{fig:bstruc_spec_fsurf}(a). In order to extract the desired information we projected the Green's function on the eigenvectors $\mathbf{c}_n(\mathbf{k})$ of the Hamiltonian of the parent compound
\begin{equation}\label{eqn:bandProjection}
 G_n(\mathbf{k},\omega) \equiv \sum_{i,j\in \mathrm{parent}} c^*_{n,i}(\mathbf{k})\left[S(\mathbf{k})\Gamma(\mathbf{k},\omega) S(\mathbf{k})\right]_{i,j} c_{n,j}(\mathbf{k})
\end{equation}
$n$ being a band index and $i,j$ local basis indices. This is only valid in the limit of small $x$ because $\Gamma$ is given in the full Hilbert space of the disordered system while
the sum in Eq.\,(\ref{eqn:bandProjection}) only runs over the subspace of the parent compound. Some of these band-projected spectral functions are shown in Fig.\,\ref{fig:bstruc_spec_fsurf}(b) and each 
of them consists of a single peak with a well defined position and spectral width even in regions where bands hybridize. The sum over these four bands (dashed gray line) essentially coincides with the total spectral function 
indicating that this decomposition works well even for 10\% substitution.\\
From this analysis we can extract the level shift of any selected band by comparing the projected spectral function of the disordered system with that of the parent compound. Concerning the broadening we assumed a Lorentzian shape
for $G_n$ which has to be evaluated at a slightly complex frequency $\omega+i\delta$
\begin{equation}\label{eqn:gfLorentz}
 G_n(\mathbf{k},\omega) = \frac{1}{\omega+i\delta - \epsilon_n(\mathbf{k}) - \Sigma_n}
\end{equation}
where $\epsilon_n(\mathbf{k})$ is the band dispersion of the parent compound, anticipating the level shifts and broadenings to be the real and imaginary part of the band self-energy $\Sigma_n$, respectively. From evaluation of  
Eq.\,(\ref{eqn:gfLorentz}) at the poles the spectral width is given by $\mathrm{Im}[1/G_n(\omega_0)]$ provided the peak position $\omega_0$ is known.\\
\indent In Fig.\,\ref{fig:bstruc_spec_fsurf}(c) a cross-section of the Fermi surface in the $(\mathbf{k}_x,\mathbf{k}_y)$ plane at $\mathbf{k}_z$=0 is shown for the parent compound - it essentially consists of three hole-like cylinders 
($\alpha$, $\beta_1$, $\beta_2$) around the center of the Brillouin zone ($\Gamma$ point) and two electron-like cylinders ($\gamma_1$, $\gamma_2$) around the zone corner (X point). By calculating level shifts and spectral broadenings
for these five bands along the $\Gamma$-Z and the X-$\Gamma$ direction, we empirically found in good approximation $\mathrm{Re}\Sigma_n\propto x$ and $\mathrm{Im}\Sigma_n\propto x$ for $x\leq0.1$. Therefore, to obtain 
general trends in a compact way, we restrict ourselves in the following to the discussion of the slopes of these quantities depending on the substituent.\\
\begin{figure}[t]
  \includegraphics[width=0.8\columnwidth]{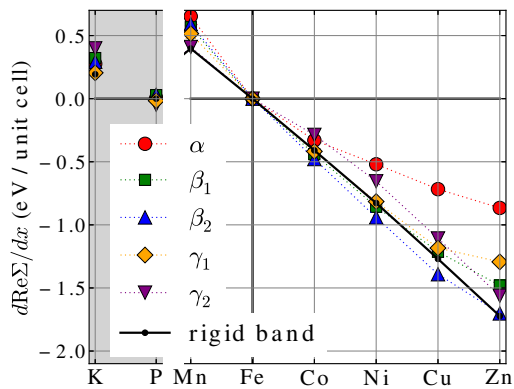}
  \caption{(color online) Slopes of level shifts $d\mathrm{Re}\Sigma_n/dx$ plotted over substituent (points) together with rigid band shift (solid line). Besides the TM substitution series for Fe, substitutions of K for Ba 
  and P for As are shown to the left (shaded)}
   \label{fig:shiftSlopes}
\end{figure}
\begin{figure}[b]
  \includegraphics[width=0.8\columnwidth]{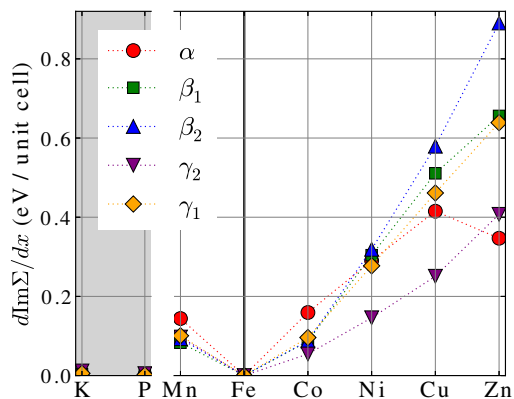}
  \caption{(color online) Slopes of the band broadenings $d\mathrm{Im}\Sigma_n/dx$ plotted over substituent. Besides the TM substitution series for Fe, substitutions of K for Ba and P for As are shown to the left (shaded)}
   \label{fig:broadSlopes}
\end{figure}
\indent In Fig.\,\ref{fig:shiftSlopes} the slopes of the level shifts $d\mathrm{Re}\Sigma_n/dx$ for the five bands are plotted for various substitutions including K for Ba and P for As. These level shifts were taken at the Fermi 
wavevector $\mathbf{k}_F$ of the disordered system. We also calculated the respective rigid band shift by integrating the density of states of the parent compound and adjusting the Fermi level to the expected number of extra 
charge carriers to be doped into the system which is shown as solid black line. In general our band selective level shifts follow the trend of the rigid band shift. The deviations from the rigid band shift get more pronounced 
with increasing difference $|\Delta_{val}|=|z_{val}(\mathrm{impurity})-z_{val}(\mathrm{parent})|$ of the valence electron number between parent compound and impurity. On the electron doped side for Ni, Cu and Zn substitution, 
electron band $\gamma_2$ and especially hole band $\alpha$ deviate more strongly from the rigid band shift than the remaining bands. These two bands have a similar orbital character - they are mainly $d_{x^2-y^2}$ bands where 
$x$ and $y$ are oriented along the projections of Fe-As bonds into the Fe planes. Overall this is consistent with the experimentally established \cite{canfield2009} dependence of the superconducting transition temperature on the 
number of extra $d$ electrons, provided that the $d_{x^2-y^2}$ bands do not significantly contribute to the pairing state.\\
\indent What, in addition, can we learn from the disorder induced lifetime effects?
Fig.\,\ref{fig:broadSlopes} shows how the slopes of the respective band broadenings $d\mathrm{Im}\Sigma_n/dx$ taken at the Fermi wavevector $\mathbf{k}_F$ of the disordered system behave under different substitutions. This 
analysis exhibits the general feature that only TM substitutions in the iron planes lead to substantial broadening effects whereas out-of-plane substitutions (Ba for K and As for P) hardly show any 
broadenings. This agrees with the fact that the major contribution to the bandstructure near the Fermi level stems from Fe $d$ states.
The TM substitutions additionally show a similar trend as already the level shifts did - an increase in $|\Delta_{val}|$ causes enhanced broadenings. This makes sense because in this context we also find an increase in the 
$l=2$ contributions to the scattering potential upon moving through the TM series. Again the broadenings are band-selective. In particular, hole bands $\beta_1$, $\beta_2$ and electron band $\gamma_1$, being of similar orbital 
character $d_{xz}$,$d_{yz}$, exhibit the same trend while among the remaining states with $d_{x^2-y^2}$ character in particular the electron band $\gamma_2$ shows considerably less broadening.\\
\indent What do these impurity scattering effects imply for superconductivity in the Ba-122 systems? Among various proposals concerning the order parameter, the most promising candidate turned out to be the $s^{+-}$-state
\cite{zhang2009,sknepnek2009} where the gap obeying $s$-wave symmetry changes sign between hole and electron pockets on the Fermi surface Fig.\,\ref{fig:bstruc_spec_fsurf}(c). The important difference  of the 
$s^{+-}$ scenario compared to conventional $s$-wave pairing is the distinction between intraband and interband scattering. While in conventional superconductors all scattering processes on nonmagnetic impurities are not
pair-breaking due to the Anderson theorem \cite{anderson1959}, interband scattering rapidly suppresses $T_c$ in an $s^{+-}$ superconductor. At the same time, intraband scattering
is irrelevant for pair-breaking but determines the residual resistivity \cite{vorontsov2009}. The impurity scattering effects we considered above reveal the joint impact of intra- and interband scattering on a band. In order to
obtain details about scattering between different bands  we consider the $T$-matrix of inserting a single impurity into the disordered crystal in the dilute limit
\begin{equation}\label{eqn:tmat}
 T_{m,n}(\mathbf{k},\mathbf{k'}) = \braket{m,\mathbf{k}}{V + V\Gamma(E_F)V +  ...}{n,\mathbf{k'}}
\end{equation}
where we used the charge self-consistent effective medium Green's function $\Gamma(E_F)$ and the difference between the impurity block and the parent block of the charge self-consistent onsite Hamiltonian for the 
impurity potential $V$. This $T$-matrix includes repeated scatterings at the same impurity up to infinite order. Via Fermi's Golden Rule we calculated the scattering rates
\begin{equation}
 w_{m,n} = 2\pi x |\mathcal{T}_{m,n}|^2 \nu_n(E_F)
\end{equation}
where $\nu_n(E_F)$ is the partial DOS of band $n$ at the Fermi level and $\mathcal{T}$ is the $\mathbf{k}$-average of $T$ over all initial and final points to get overall trends. We additionally averaged over the two outer hole 
bands $\langle\beta_1,\beta_2\rangle\rightarrow\beta$ and over the two electron bands $\langle\gamma_1,\gamma_2\rangle\rightarrow\gamma$ because their $\mathbf{k}$-space anisotropy mutually cancels. The results divided by 
concentration $x$ for these intraband $\mathcal{V}$ and interband $\mathcal{U}$ scattering rates between the effective three bands are shown in Fig.\,\ref{fig:tmat}.
We find that the intraband scattering rates inside the hole bands are the largest whereas the electron intraband and electron-hole interband scattering rates are smaller by a factor of two. The interband
scattering between hole bands $\alpha$ and $\beta$ is negligible. This behavior is universal for all substitutions and has two important implications: first it indicates that an $s^{+-}$ state can exist in these 
systems together with considerable impurity scattering because the latter is dominated by intraband scattering which is not relevant for pair breaking. Secondly this 
suggests that the transport properties, being primarily an intraband scattering phenomenon, are mainly governed by the hole bands which for Co substitution was indeed pointed out in transport experiments \cite{rullier2009}.\\
Furthermore the scattering rates show the same trend as already the band broadenings did: they grow for increasing $|\Delta_{val}|$ and are only relevant for TM substitution.
Under the assumption of $s^{+-}$ superconductivity this tells us that Co substitution is more strongly pair-breaking than K, and for TM substitution the pair-breaking strength increases with $\Delta_{val}$. Clearly this can be 
connected with experimental facts because at optimal doping for the critical temperatures among the different substitutions it holds $T_c($K$)>T_c($Co$)>T_c($Ni$)>T_c($Cu$)$. Mn substitution does not follow this trend due to 
our above mentioned non spin-polarized calculations which shows the importance of a local magnetic moment for impurity scattering in case of Mn in contrast to the other substitutions.\\
\indent In conclusion, we have presented a systematic \emph{ab-initio} study of disorder effects for a variety of substitutions in the Ba-122 compound. 
For TM substitution, shifts of individual bands at the Fermi level are, to first approximation, compatible with the picture of adding charge carriers to the system. However there are deviations from this rigid band behavior which,
like the spectral broadenings, are connected with the orbital composition of the band and the substitutional site. An analysis distinguishing intra- and interband scattering in the dilute limit substantiates the robustness of 
the proposed $s^{+-}$ pairing state. Among different substitutions this analysis indicates growing pair-breaking with increasing number of $d$ electrons for TM substitution in qualitative agreement with experiment.\\
We acknowledge R. Eder, K. Grube, F. Hardy, M. Hoyer and H. v. L\"ohneysen for fruitful discussions and corrections. All plots within this Letter have been generated by matplotlib \cite{hunter2007}, an open source project.

\bibliography{bfaShiftsBroadenings}

\end{document}